\documentclass[aps,prl,reprint,amssymb,showpacs,superscriptaddress,notitlepage,onecolumn]{revtex4-2}
\usepackage{bm}
\usepackage{graphicx}
\usepackage{hyperref}
\usepackage{dcolumn}
\usepackage{braket}
\usepackage{natbib}
\usepackage{amsmath}
\usepackage{color}
\usepackage{mathtools,amssymb}
\usepackage{soul}
\usepackage{float}
\usepackage{amssymb}
\usepackage{esint}
\usepackage{mathtools}
\usepackage{physics}
\usepackage{makecell}
\usepackage{multirow}
\usepackage{array}
\usepackage{geometry}
\usepackage{url}
\usepackage{adjustbox}
\begin{document}
\title{Dynamic control of 2D non-Hermitian photonic corner states in synthetic dimensions}

\author{Xinyuan Zheng}\email{xzheng16@terpmail.umd.edu}
\altaffiliation{Equally contributing authors}
\affiliation{Institute for Research in Electronics and Applied Physics, University of Maryland, College Park, Maryland 20742, USA}

\author{Mahmoud Jalali Mehrabad}\email{mjalalim@umd.edu}
\altaffiliation{Equally contributing authors}
\affiliation{Joint Quantum Institute, University of Maryland, College Park, MD 20742, USA}

\author{Jonathan Vannucci}
\affiliation{Joint Quantum Institute, University of Maryland, College Park, MD 20742, USA}

\author{Kevin Li}
\affiliation{Joint Quantum Institute, University of Maryland, College Park, MD 20742, USA}

\author{Avik Dutt}
\affiliation{Department of Mechanical Engineering, and Institute for Physical Science and Technology, University of Maryland, College Park, Maryland 20742, USA}

\author{Mohammad Hafezi}
\affiliation{Joint Quantum Institute, University of Maryland, College Park, MD 20742, USA}

\author{Sunil Mittal}\email{s.mittal@northeastern.edu}
\affiliation{Department of Electrical and Computer Engineering, Northeastern University, Boston, MA, USA}

\author{Edo Waks}\email{edowaks@umd.edu}
\affiliation{Institute for Research in Electronics and Applied Physics, University of Maryland, College Park, Maryland 20742, USA}

%%%%%%%%%%%%%%%%%%%%%%%%%%%%%%%%%%%%%%%%%%%%%%%%%%%%%%%%%%%%%%%%%%%%%%%%%%%%%%%%%%%%%%%%%%%%%%%%%%%%%%%%%%%%%%%%%%%%%%%%%%%%%%

\begin{abstract}

Non-Hermitian models describe the physics of ubiquitous open systems with gain and loss. One intriguing aspect of non-Hermitian models is their inherent topology that can produce intriguing boundary phenomena like resilient higher-order topological insulators (HOTIs) and non-Hermitian skin effects (NHSE). Recently, time-multiplexed lattices in synthetic dimensions have emerged as a versatile platform for the investigation of these effects free of geometric restrictions. Despite holding broad applications, studies of these effects have been limited to static cases so far, and full dynamical control over the non-Hermitian effects has remained elusive. Here, we demonstrate the emergence of topological non-Hermitian corner states with remarkable temporal controllability and robustness in a two-dimensional photonic synthetic time lattice. Specifically, we showcase various dynamic control mechanisms for light confinement and flow, including spatial mode tapering, sequential non-Hermiticity on-off switching, dynamical corner state relocation, and light steering. Moreover, we establish the corner state's robustness in the presence of intensity modulation randomness and quantitatively determine its breakdown regime. Our findings extend non-Hermitian and topological photonic effects into higher synthetic dimensions, offering remarkable flexibility and real-time control possibilities. This opens avenues for topological classification, quantum walk simulations of many-body dynamics, and robust Floquet engineering, free from the limitations of physical geometries.
\end{abstract}

%%%%%%%%%%%%%%%%%%%%%%%%%%%%%%%%%%%%%%%%%%%%%%%%%%%%%%%%%%%%%%%%%%%%%%%%%%%%%%%%%%%%%%%

\maketitle

\section{Introduction}

Non-Hermitian systems host a range of intriguing phenomena in physics, such as reconfigurable light routing \cite{zhao2019non}, potential for enhanced sensitivity \cite{chen2017exceptional,wiersig2020review} and unidirectional invisibility \cite{feng2013experimental}, that are deeply rooted in symmetry and topology. One such phenomenon is the non-Hermitian skin effect (NHSE) where a macroscopic fraction of the eigenmodes of a finite system become exponentially localized at its boundary \cite{nasari2023non,lin2023topological}. This localization is a direct consequence of the nontrivial (topological) winding of the system's eigenvalues in the complex energy plane \cite{yao2018edge,kawabata2018anomalous,gong2018topological}. Disorder and small variations in the system do not change the winding number which is a topological invariant \cite{gong2018topological}.

Over the last few years, the NHSE has been demonstrated on a variety of platforms \cite{nasari2023non,zhang2022review,okuma2023non}. Exemplary platforms include acoustics and phononics \cite{zhou2023observation}, topo-electric circuits \cite{zou2021observation}, and photonics \cite{weidemann2020topological}. These developments are in part motivated by the profound impact of NHSE on band topology \cite{yao2018edge,yokomizo2019non,wang2021topological}, spectral symmetry \cite{xiao2021observation}, and dynamics \cite{wang2021simulating,liang2022dynamic}. Particularly in photonics, recently the NHSE has enabled intriguing demonstrations of the tuneable directional flow of light \cite{lin2023manipulating}, near-field beam steering \cite{liu2022complex}, engineering arbitrary band topology \cite{wang2021generating} and topological funneling of light \cite{weidemann2020topological}. Nevertheless, these demonstrations have been limited to systems that can be effectively described by time-independent Hamiltonians \cite{chalabi2019synthetic}. The introduction of time-dependent non-Hermitian Hamiltonians can lead to a dynamic control over the skin effect and also lead to fundamental advances in novel non-Hermitian topological phases that are not accessible using time-independent systems.

Here we demonstrate dynamical control of the two-dimensional non-Hermitian photonic skin effect, that is, corner states, using purely synthetic temporal dimensions. Specifically, using time-multiplexed light pulses in fiber loops, we show manipulation of the gain/loss in the system at a scale that is faster than the dynamics of light pulses in the lattice. Using this dynamical manipulation, we demonstrate adiabatic control over the degree of localization of the corner states, adiabatic tweezing of light where we move the corner states along a predefined trajectory in the lattice, and 2D funneling of light where photons always funnel to the corner state irrespective of their initial position in the 2D lattice.  Finally, by introducing controlled disorder in the system in the form of random variations in gain and loss, we quantitatively investigate the robustness of the corner states against such disorders. Our work opens up an avenue to explore the rich physics of time-dependent non-Hermitian models such as non-Hermitian Floquet systems.

\section{2D quantum walk with non-Hermitian topology}

\begin{figure}
    \centering    
    \includegraphics[width=\columnwidth]{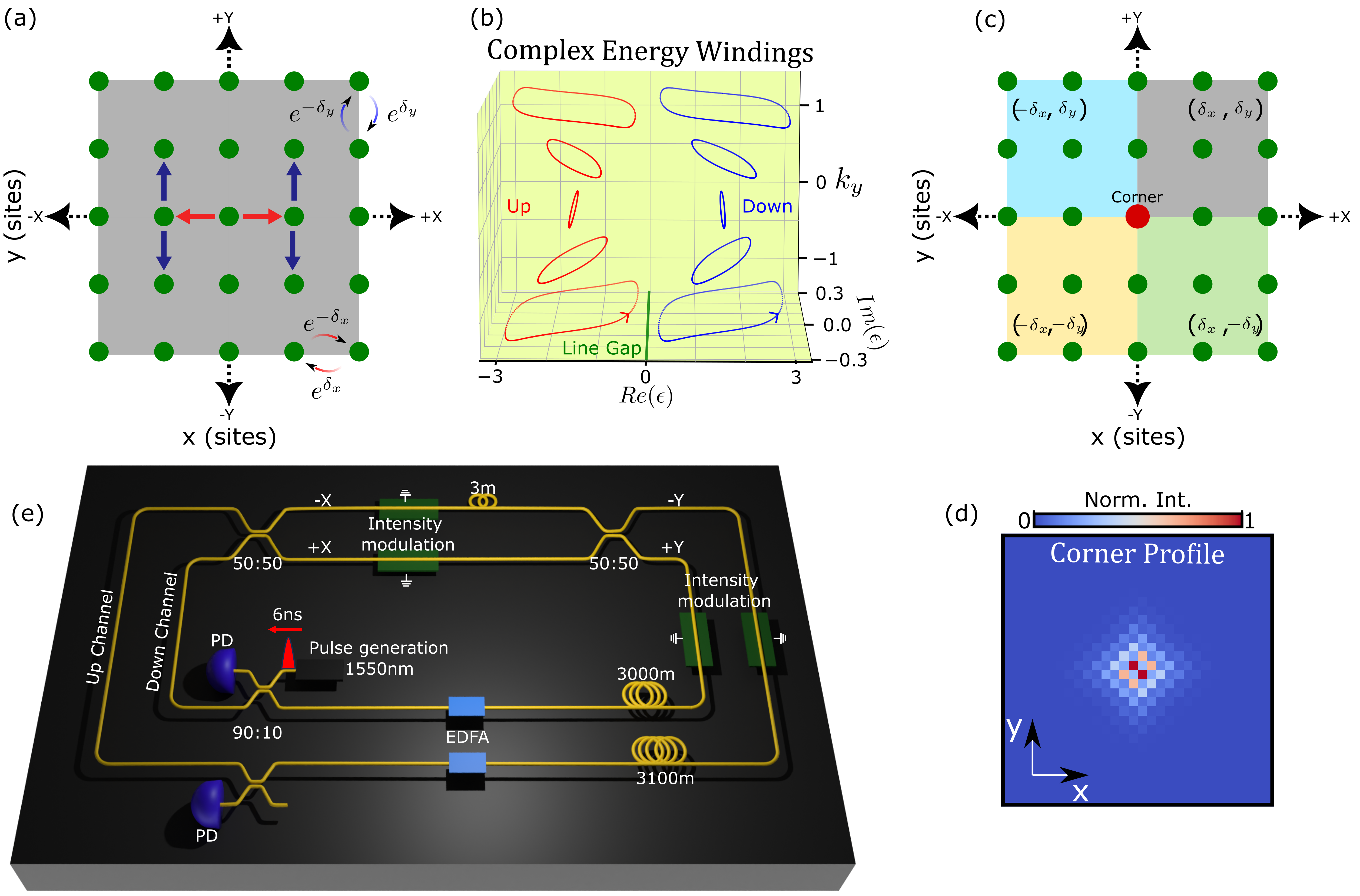}
    \caption{\textbf{Two-dimensional quantum walk in non-Hermitian synthetic lattice.} (a) Example of photonic quantum walk in a 2D synthetic lattice. The blue and red curved arrows show the direction-dependent loss and gain.  (b) Winding of effective eigenenergies $\epsilon_{up/down}(k_x,k_y)$ in the complex energy plane for a single bulk non-Hermitian lattice with periodic boundary condition, showing line-gapped topology (indicated by the green line). Here we choose five different values $k_y \equiv 0, \pm \pi/4, \pm 3\pi/8$. (c) Four bulk lattices with different gain-loss patterns are glued along their edges to form a corner. Note that $\delta_x>0$ implies gain for a step towards $-X$ and loss for a step toward $+X$. For $\delta_x<0$ the gain-loss is inverted. A similar rule applies for $\delta_y$. 
    (d) Averaged spatial profile of corner states formed in the system shown in (c), by taking non-Hermitian parameters $\delta_x=\delta_y=0.175$. The lattice size is 30$\times$30. (e) The time-multiplexed experimental scheme, with which the lattice parameters can be (dynamically) controlled by the intensity modulators.}
    \label{schematic}
\end{figure}
Our system simulates a discrete-time quantum walk of photons on a two-dimensional non-Hermitian square lattice, as illustrated schematically in Figure \ref{schematic}(a). Specifically, we implement a split-step walk where the walker first randomly steps to either left or right (corresponding to the $X$ direction) with equal probability, then up or down (corresponding to the $Y$ direction). To introduce non-Hermiticity, we introduce an additional gain $e^{+\delta_x}$ when the walker steps to the left, and an additional loss $e^{-\delta_x}$ when the walker steps to the right. Similarly, the walker experiences a gain $e^{+\delta_y}$ when moving down and a loss of $e^{-\delta_y}$ when moving up. The parameters $\delta_x$ and $\delta_y$ then indicate the degree of non-Hermiticity of the walk.

For this quantum walk, a concept of complex energy can analogously be defined, by solving for the eigenmodes of the non-unitary quantum walk evolution operator $\hat{U}$ and taking the logarithm of the corresponding eigenvalue $u_j$. Namely, this can be formulated as $\epsilon_j = ilog(u_j)$, where $\hat{U}\ket{u_j} = u_j\ket{u_j}$. If we further impose periodic boundary conditions (PBCs) in both $x$ and $y$ for the bulk in Figure \ref{schematic}(a), we can apply the Bloch theorem for the walk and obtain the complex energy bands $\epsilon_{up/down}(k_x,k_y)$. (The two bands seen in Figure \ref{schematic}(b) arise due to the up/down channel configuration of our experiment, see supplementary information (SI) for derivation details). The non-unitary time evolution of the walk leads to a non-trivial winding of $\epsilon(k_x,k_y)$ for each band in the complex energy plane as one continuously varies Bloch vector $(k_x,k_y)$ along a certain curve in the Brillouin zone. To illustrate this, in Figure \ref{schematic}(b), we plot the complex energies $\epsilon_{up/down}(k_x,k_y)$ of the bulk lattice shown in Figure \ref{schematic}(a) as we vary $k_x$ from $-\pi$ to $\pi$ while keeping $k_y$ fixed to different values $0,\pm \pi/4,\pm 3\pi/8$. As $(k_x,k_y)$ varies along each of these directed horizontal curves in the Brillouin zone, both $\epsilon_{up}(k_x,k_y)$ and $\epsilon_{down}(k_x,k_y)$ winds one loop in the counterclockwise direction, thus exhibiting an integer-valued winding number $-1$. This is a topological invariant for our non-Hermitian quantum walk. Also, the two winding loops contributed from the two bands $\epsilon_{up/down}$ exhibit a line-gapped topology \cite{okuma2020topological}, such that the two winding loops never cross the line $Re(\epsilon)=0$ in the complex plane. Windings of complex energy along other curves in the Brillouin zone are shown in the SI.

In a finite system, the non-trivial winding of the complex energies and the associated 2D non-Hermitian skin effect\cite{okuma2020topological} is manifested as corner states, that is, localization of the walker can happen at an interface between regions with opposite windings (or bulk band topologies). Figure \ref{schematic}(c) shows one exemplary case which consists of four distinct regions, represented by the four different color patches. The gray patch is identical to the system described in Figure \ref{schematic}(a). The other three regions exhibit an inverted gain-loss relation (indicated by a change in the sign of the gain parameter) either along the $x$ or $y$-axis, or both. This inversion of gain-loss leads to different windings for each region. Non-Hermitian skin effect occurs in such a system, and we numerically verify in Figure \ref{schematic}(d) that the averaged eigenmodes of the quantum walk exhibit clustering at the junction between the four regions - as indicated by the red dot in Figure \ref{schematic}(c).

To simulate the quantum walk described above, we use classical light pulses in a time-multiplexed setup shown in Figure \ref{schematic}(e). We note that for this linear system, the evolution of classical light pulses in the lattice exactly follows that of the quantum walk of single photons in the lattice. We map the state space of the 2D square lattice into different time-delays in two fiber feedback loops, as introduced in previous works \cite{schreiber20122d,nitsche2016quantum}. To introduce non-Hermiticity, we use four intensity modulators that introduce individually controllable loss when the walker moves along any direction. We also use two erbium-doped fiber amplifiers (EDFAs) that provide gain in the system, and together with the intensity modulators, introduce a gain-loss mechanism that can be controlled at each step of the walker. We specifically choose electro-optic modulators with a high bandwidth to allow reconfigurability of the system's topology at each step of the quantum walk. A full discussion of the experimental setup is provided in the SI.

\begin{figure}
    \centering    
    \includegraphics[width=\columnwidth]{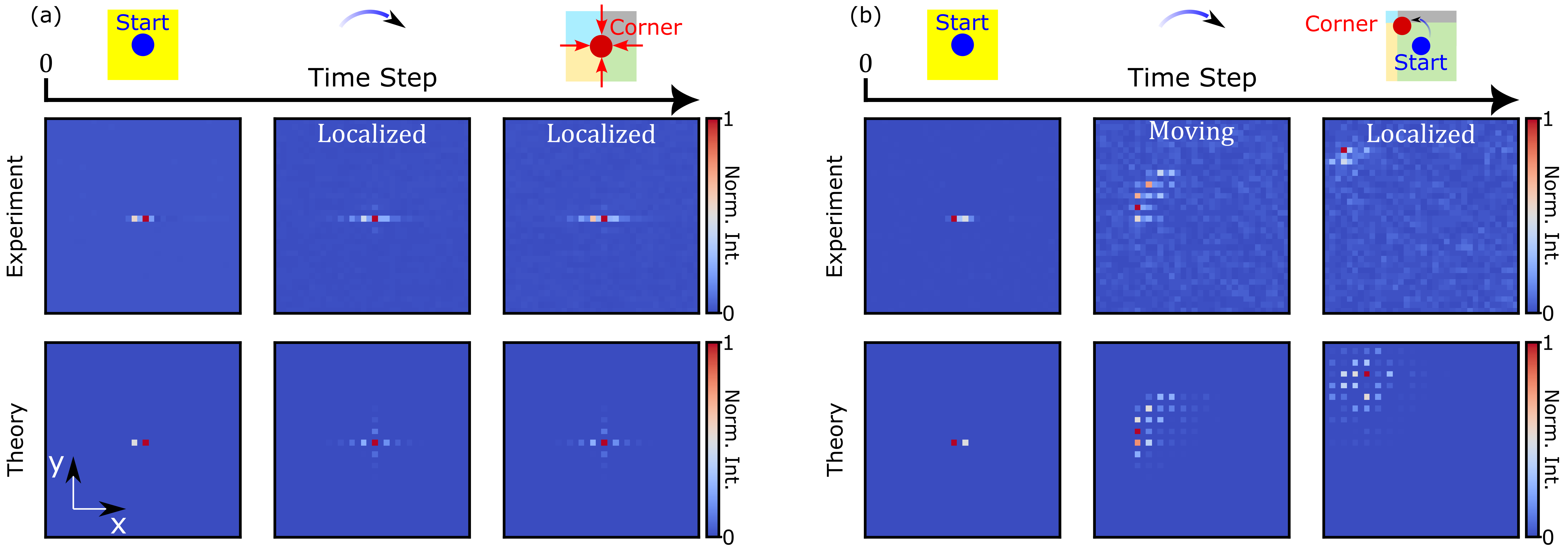}
    \caption{\textbf{Light localization and light funneling for static control.} (a) Light localization at the corner state located at $(x,y)=(0,0)$ for non-Hermitian parameter $|\delta_x|=|\delta_y|=0.175$. Here a single pulse is initialized at $(x,y)=(0,0)$ in the up channel.
    (b) Light funneling for the same lattice parameter and pulse initialization, but the corner state is located at $(x,y)=(-10,10)$. Here the skin effect allows light to flow to the corner state and localize there. In both (a) and (b), from left to right the snapshots are shown for time steps $1,9,21$, respectively.}
    \label{static}
\end{figure}

\section{Results}
\section{Skin effect under static control}
To show the presence of non-Hermitian corner states, we first construct the model as shown in Figure \ref{schematic}(c), with the corner located at the lattice origin $(x=0,y=0)$. We inject a single light pulse into the time bin corresponding to the lattice origin and choose non-Hermitian parameters $|\delta_x|$ and $|\delta_y|$ to be 0.175 as in Figure \ref{schematic}(c). In Figure \ref{static}(a), we plot the snapshots of the light distribution in the lattice for different time steps $1$, $9$, and $21$, which are obtained by measuring the pulse power at each time bin. The evolution of distribution shows that the walker stays localized at the origin, confirming the presence of a corner state. In sharp contrast, when we set $\delta_x$ and $\delta_y$ to 0, we observe a significant spreading of the intensity distribution, indicating the absence of any corner states (see the SI for the experimental data).  

Having shown the localization of light at the corner state, we next demonstrate the skin-effect-induced funneling of light. Namely, the system dynamics bring any initial state towards the corner states. We set the corner state to be at the lattice site $(x=-10,y=10)$ while light pulses are still injected at $x=0,y=0$, which is now in the bulk of the lattice (Figure \ref{static}(b)). As the system evolves, initially light spreads in bulk, but finally converges to the corner site. As shown in the SI for several different lattice configurations, light pulses always converge to the corner regardless of the initialization location. This funneling of light to the corner state is a manifestation of the skin effect where all the eigenmodes of the system are localized at the corner. Our experimental results are in good agreement with our theoretical prediction shown in Figure \ref{static}(b).

\begin{figure}
    \centering    
    \includegraphics[width=0.8\columnwidth]{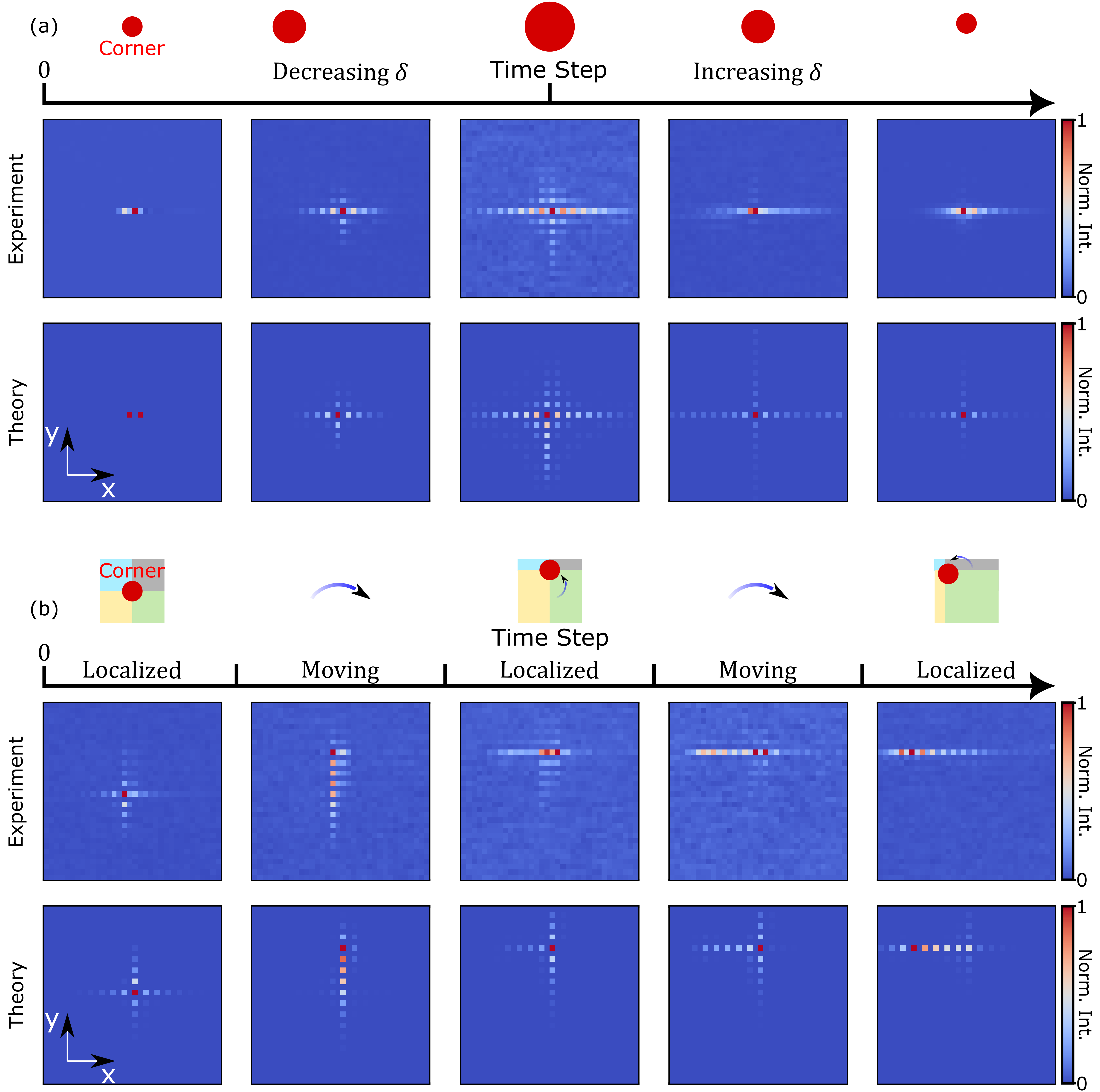}
    \caption{\textbf{Dynamical control of the corner state}. (a) Dynamical control of corner state spatial profile. As the non-Hermitian parameter is adiabatically reduced from $|\delta_{x,max}|=|\delta_{y, max}|=0.175$ to $|\delta_{x,max}|=|\delta_{y, max}|=0.02$ and back to $|\delta_{x,max}|=|\delta_{y, max}|=0.175$, the corner state becomes delocalized and then localized. From left to right the snapshots are shown for time steps $1,9,17,25,37$, respectively.
    (b) Dynamically tweezing localized light along a designed “L”-shaped trajectory using the skin effect. Localized light is first moved in the $+Y$ direction for 8 steps and then to the $-X$ direction for 10 steps.
    }
    \label{dynamical}
\end{figure}

\section{Dynamically controlling the non-Hermitian lattice and skin effect}

The use of time as a synthetic dimension allows us to dynamically reconfigure our non-Hermitian lattice as a function of time. Specifically, by controlling the intensity modulators at each time step of the quantum walk, we achieve temporal modulation of the gain/loss parameters $\delta_x(t)$ and $\delta_y(t)$ such that they are time-dependent. Using this time dependence, first, we demonstrate dynamical control over the degree of localization of the corner states. At the start of the evolution, we adopt the configuration as in Figure \ref{schematic}(c) and set $|\delta_x(0)|=|\delta_y(0)|=0.175$, and inject a single light pulse at the corner state situated at the origin. As the system evolves, we reduce both $|\delta_x|, |\delta_y|$ by $50\%$ for every four time-steps and continue doing so until step $16$ (Figure \ref{dynamical}(a)). Because of this reduction, we observe that the corner states become less confined to the origin. This is because the smaller non-Hermitian parameter exhibits eigenmodes distributed over a larger area, as predicted theoretically (see the SI). Thereafter, starting from step $17$, we reverse the process, that is, we increase the gain /loss parameters $|\delta_x|, |\delta_y|$ back to its original value at the same rate. We now observe a relocalization of light at the origin.

Next, we demonstrate adiabatic repositioning of the corner states in the lattice. We use the same lattice geometry shown in Figure \ref{schematic}(c) and fix the non-Hermitian parameter to $\delta_x=\delta_y=0.175$. As the system evolves, we adiabatically move the interface between the four distinct topological regions, repositioning the corner state as a function of time. We first move the position of the corner state upwards for 8 unit cells, and then leftward for 10 unit cells. As before, we inject light pulses at the corner state. As the system evolves, we observe that the center of the intensity distribution follows the position of the corner state as it adiabatically moves along the given "L"-shaped trajectory from its initial location $(x=0,y=0)$ to its final location at $(x=-10,y=8)$. Furthermore, during this process, the intensity distribution remains tightly localized close to the corner state. Evidently, the corner serves as a non-Hermitian tweezer of light, which allows us to adiabatically move trapped photons along a given trajectory in the synthetic lattice. Note that non-Hermitian light steering has been demonstrated in real-space lattices \cite{zhao2019non}, and our demonstration in synthetic time dimensions portends the potential for such photonic control using the temporal degree of freedom of light.

\begin{figure}
    \centering    
    \includegraphics[width=\columnwidth]{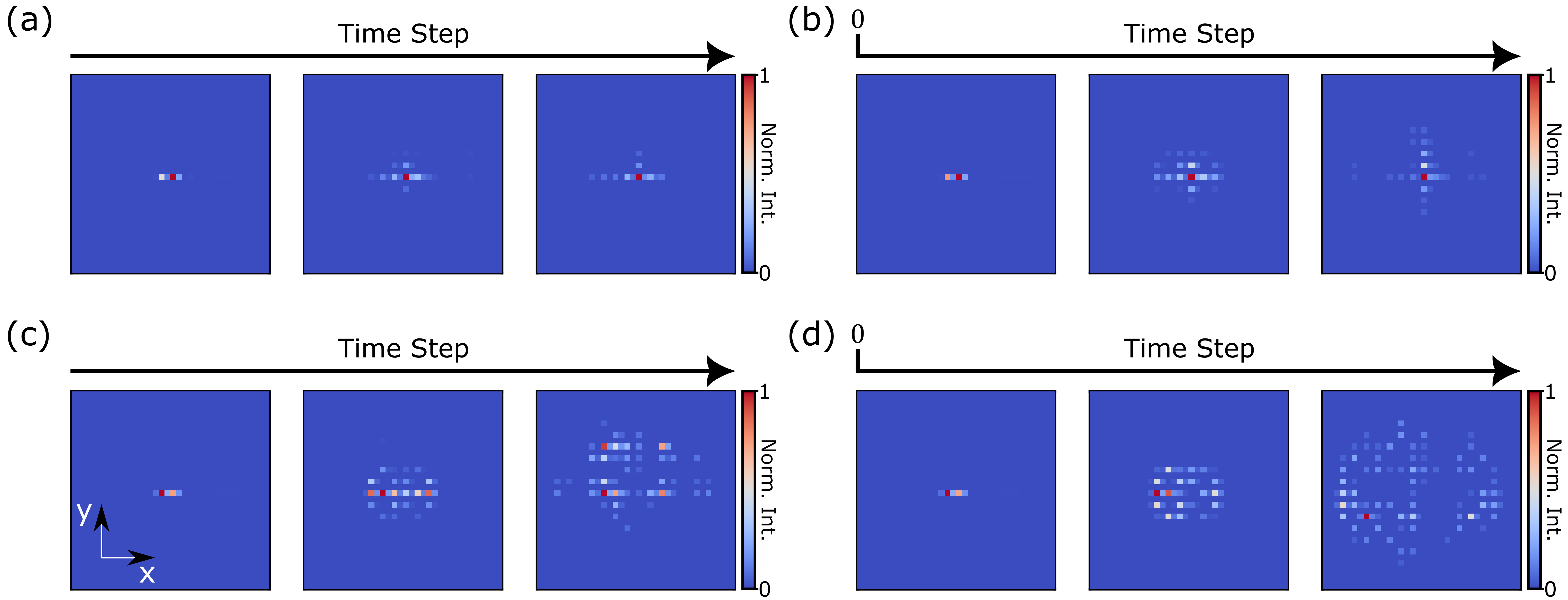}
    \caption{\textbf{Robustness of the skin effect at the presence of different degrees of lattice disorder.} The randomness is introduced to the intensity modulation of the lattice and the pulse is injected at $(x,y)=(0,0)$. (a-b) Experimental observation of robustness of the corner state and skin effect in a lattice with moderate disorder $(\eta_y=\eta_x = 0.5,1)$. Here the disorder leads to a relaxed spatial confinement of light without breaking the localization of light. (c-d) Breakdown of the localization in the presence of strong disorder $(\eta_y=\eta_x = 1.5, 2)$. Light can diffuse arbitrarily far away from the original location. In (a-d), from left to right the snapshots are shown for time steps $1,5,13$, respectively.
    }
    \label{robustness_Evolution}
\end{figure}

\section{Robustness of the skin effect}

The topological nature of the non-Hermitian skin effect ensures its robustness against disorder in gain/loss parameters $\delta_x, \delta_y$. To quantitatively investigate this robustness, we introduce a disorder on the gain/loss. At each lattice site, we randomly pick both $\delta_x, \delta_y$ from a uniform distribution on the interval $[ \delta_{max}(1-\eta), \delta_{max}]$, where max is the maximum gain parameter and is the disorder parameter which quantifies the variance of the gain parameter. In our experiment, we vary the disorder parameter between 0 (no disorder) and 2 (max disorder).

\begin{figure}
    \centering    
    \includegraphics[width=\columnwidth]{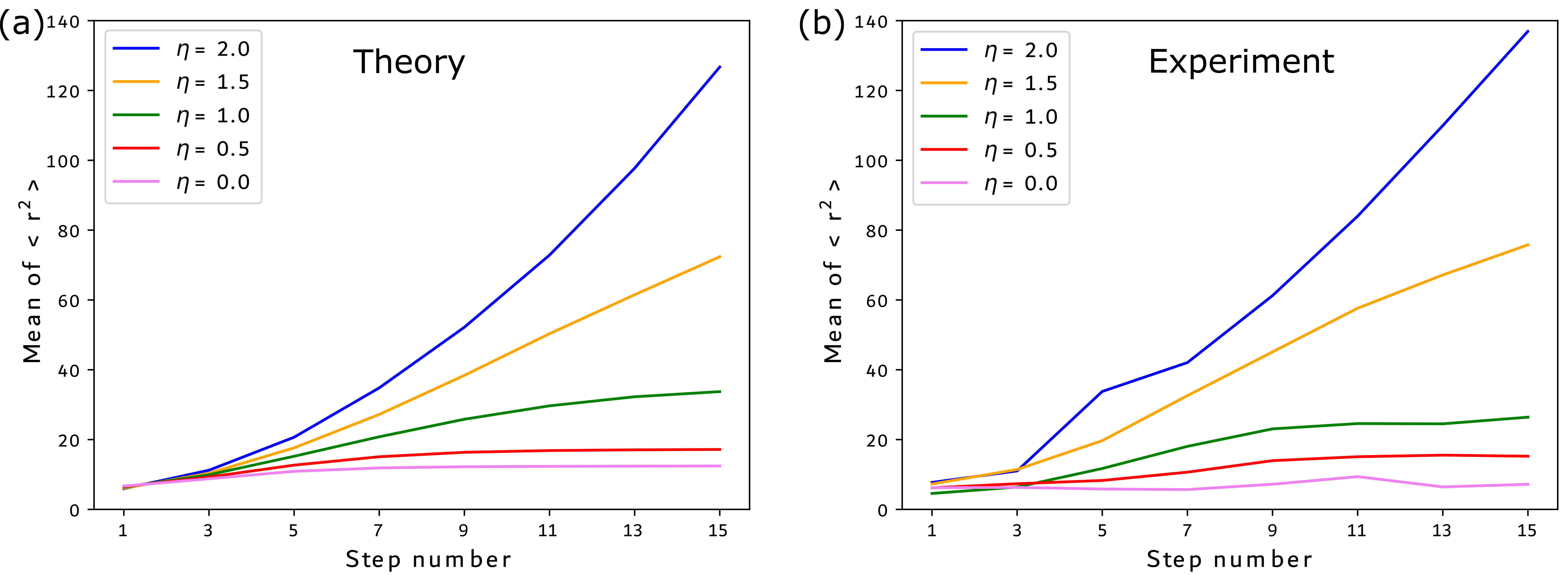}
    \caption{\textbf{(a) Theory and (b) Experimental evolution of space-averaged displacement $<r^2>=<x^2+y^2>$ as a function of step number under different lattice disorders.} For disorder strength lower than the threshold $\eta=1$, the disorder increases the effective spatial diameter of the corner state, as shown in the evolution of average displacement with time. For disorders higher than the threshold, light diffusively spreads to large distances on the lattice.
    }
    \label{robustness_Trejectories}
\end{figure}

We find that the skin effect is robust when the disorder parameter $\eta<1$. In Figure \ref{robustness_Evolution}(a-b), we plot the evolution of light pulses in the lattice for two different values $\eta=0.5$ and $\eta=1$. For both cases, we inject light pulses at the corner state located at the origin. We observe that, even though the localization of the intensity distribution reduces as disorder increases, the distribution still stays confined around the origin, indicating the existence of corner states even in the presence of disorder. Nevertheless, once we increase the disorder parameter to 1.5 and 2 (Figure \ref{robustness_Evolution}(c-d)), the intensity distribution diffusively spreads away from the origin, indicating the breakdown of corner states. Our experimental observation agrees with the intuitively expected behavior that, for $\eta<1$, even though there is a disorder in the modulation amplitudes, the gain for the step towards the corner is always larger than that of the outwards direction. Thus the four regions maintain their distinct non-Hermiticity and the corner state exists. But, when $\eta>1$, a direction-dependent gain for the time steps is no longer always valid and therefore the four regions are no longer distinct and the corner state ceases to exist. 

To better characterize the robustness and breakdown of the skin effect, we compute the evolution of the mean-square displacement of the intensity distribution in the lattice as a function of time. The mean-square displacement is quantified as $<r^2>(n)=\sum_{x,y}P_{x,y}(n)(x^2+y^2)$, where $P_{x,y}(n)$ is the time-varying intensity distribution of light. Figure \ref{robustness_Trejectories} shows the calculated $<r^2>(n)$ for several values of the disorder parameter for both theoretical calculations and experimentally measured values. Each experimental curve corresponds to an average of four independent experimental realizations of disorder, while the theory corresponds to four averages. 
Due to the limited size of the lattice $30\times30$, we only collect data from step $1$ to step $15$, and plot $<r^2>(n)$ for the odd steps. The violet, red, and green curves correspond to the weak disorder, with disorder parameters being $0$, $0.5$, and $1$, respectively. All three curves saturate to a fixed value which is well below a certain threshold. This behavior thus implies the robustness of the skin effect. But for larger disorders of $1.5$ and $2$, corresponding to the yellow and blue curve, the mean squared distance does not converge. Instead, it spreads out until it becomes limited by the finite size of the lattice, indicating a complete breakdown of the skin effect.

\section{Summary and outlook}
In conclusion, we demonstrated robust dynamical control over the photonic non-Hermitian skin effect in a 2D synthetic lattice. We created a corner state that localizes light and dynamically tuned the degree of light localization. Moreover, we dynamically steered trapped light along any given trajectory in the 2D lattice. We also demonstrate the robustness of the skin effect under lattice disorder below a certain threshold.

Looking forward, the dynamic techniques developed in this work can be further applied to investigate Floquet non-Hermitian models \cite{weidemann2022topological,zhou2023non}, in particular in synthetic dimensions \cite{yuan2018synthetic}. Further, one can create an analogue of on-site interaction by imposing a nonlinear phase shifter after the linear optical transformations, and investigate non-Hermitian models of interacting particles \cite{gliozzi2024many}. Such nonlinearities could also have implications in the recently discovered regime of topological frequency combs \cite{mittal2021topological,jalali2023topological,flower2024observation} due to the periodic temporal pulses that define our platform. Moreover, the two-fold spin characteristics in our system can potentially be extended to non-Hermitian models for lattice gauge theories with higher spins and non-Abelian statistics, by increasing the number of loops \cite{pang2024synthetic}. Finally, our non-hermitian lattice can be enriched with engineered synthetic gauge fields as demonstrated recently for both Hermitian \cite{chalabi2020guiding} and non-Hermitian models\cite{lin2023manipulating}, to explore intriguing proposals such as the quantum Skin Hall effect \cite{ma2023quantum}.

%%%%%%%%%%%%%%%%%%%%%%%%%%%%%%%%%%%%%%%%%%%%%%%%%%%%%%%%%%%%%%%%%%%%%%%%%%%%%%%%%%%%%%%%%%%%%%%%%%%%%%%%%%%%%%%%%%%%%%%%%%%%%%

\section{Acknowledgements}
The authors wish to acknowledge fruitful discussions with Shanhui Fan and Nathan Schine. This research was supported by The Office of Naval Research ONR-MURI grant N00014-20-1-2325, AFOSR FA95502010223, NSF PHY1820938, NSF DMR-2019444, the Army Research Laboratory grant W911NF1920181 and EPSRC grant EP/V026496/1.

%%%%%%%%%%%%%%%%%%%%%%%%%%%%%%%%%%%%%%%%%%%%%%%%%%%%%%%%%%%%%%%%%%%%%%%%%%%%%%%%%%%%%%%%%%%%%%%%%%%%%%%%%%%%%%%%%%%%%%%%%%%%%%
\bibliography{Main.bib}
\newpage

\newpage
%\onecolumngrid

%%%%%%%%%% Merge with supplemental materials %%%%%%%%%%
%%%%%%%%%% Prefix a "S" to all equations, figures, tables and reset the counter %%%%%%%%%%
\setcounter{equation}{0}
\setcounter{figure}{0}
\setcounter{table}{0}
\setcounter{page}{1}
\makeatletter
\renewcommand{\theequation}{S\arabic{equation}}
\renewcommand{\thefigure}{S\arabic{figure}}
\pagenumbering{roman}

\section{Supplementary Information: Dynamic control of 2D non-Hermitian photonic corner states in synthetic dimensions}
\subsection{Experimental Setup}
Here we show the details of the experimental setup.

\begin{figure}[b]
%\subfloat[\label{sfig:1}]{}
\includegraphics[width=0.9\columnwidth]{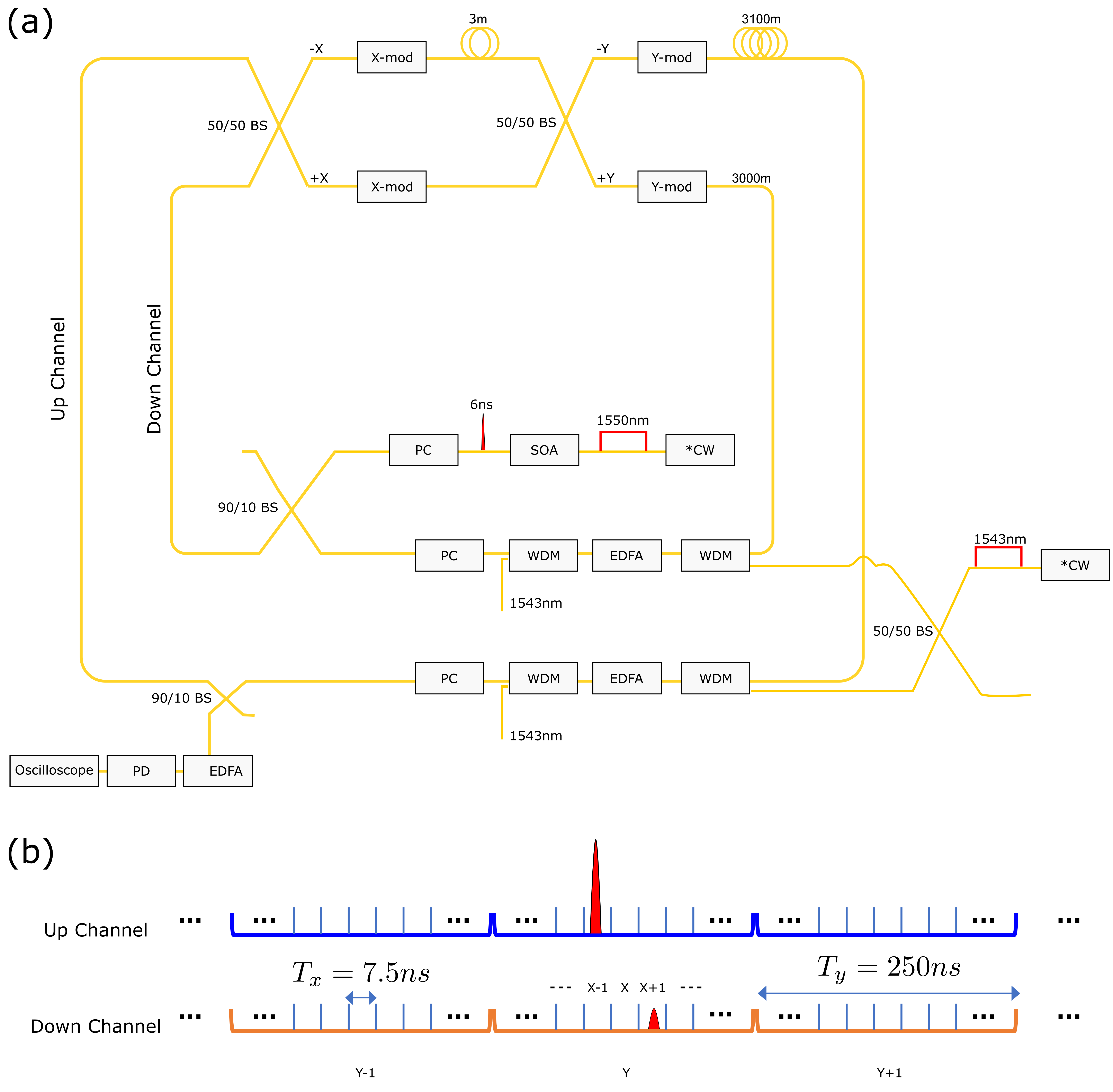}
\caption{\label{fig:s1} \textbf{Sketch of the complete experimental setup and encoding scheme.} (a) Details of the experiment, with continuous wave (CW) laser, polarization control (PC), wavelength division multiplexer (WDM), photodiode (PD), beam splitter (BS), intensity modulators (X/Y-mod), erbium doped fiber amplifier (EDFA) and semiconductor optical amplifier (SOA). (b) Encoding the 2D lattice in time bins in two fiber loops. The two loops are named ``up channel" and ``down channel", respectively.}
\end{figure}

To encode the 2D lattice in time we consider two fiber loops shown in Figure \ref{fig:s1}(a), labeled up channel and down channel. The length of each fiber loop is $\sim$ 3 (km), and one circulation of light in the loop is equivalent to one step of the walk. Hence, we can encode the entire 2D lattice within a time-duration (or time-delay) of $\sim$ 15000 (ns) without mixing time-bins in step $n$ and step $(n+1)$. As in Figure \ref{fig:s1}(b), we first encode $30$ ``Y"-time bins in both the up and down channel, each of time duration 250 (ns) in a total time duration of 7500 (ns). Each ``Y"-time bin is then occupied by $30$ ``X"-time bins, each of time duration 7.5 (ns). At any time, the state of the system is thus represented by a complex vector $(U_{x,y},D_{x,y})$, encoded in the phase and amplitude of the light pulse circulating in the two fiber loops. 

To initialize the system, we inject a single pulse into the down channel of the fiber loop. We use a continuous wave CW laser with 1550 (nm) wavelength (Optilab DFB-1551-SM-10) and by modulation of this laser using a Thorlabs SOA (SOA1013SXS), we have generated pulses of width $\sim$ 6 (ns) at a repetition rate of 1 (pulse/ms). We then control the polarization of the laser with an inline fiber polarization control (PC) before injecting the light into the down channel with a $90/10$ beam splitter. Note that we use two identical $90/10$ beam splitter, one for each channel. The $90/10$ beam splitter in the down channel is used to inject light into the quantum walk, whereas the $90/10$ beam splitter in the up channel is used to weakly couple light pulses out of the quantum walk so that we can measure the pulse power after $n$ steps of evolution. 

As a pulse enters the system, by default we recognize it as entering the $(x=0,y=0)$ time bin, and thus the initial state is $D_{0,0}=1$. The pulse then sequentially passes through a $50/50$ beam splitter denoted as $\pm X$-beam splitter, a pair of time-varying intensity modulators (Optilab IMP-1550-20-PM) is used to impose the correct gain/loss as each time bin $(x,y)$ passes through it, controlled by RF signal generated from Teledyne Lecroy arbitrary waveform generator (T3AWG3252). We then impose a delay of 3 (m) in the up channel and no delay in the down channel. The same procedure then repeats for $Y$, as shown in Figure \ref{fig:s1}(a), except the difference in delay between the up and down channels is 100 (m).

To combat photon loss in the walk, we use two Thorlabs erbium-doped fiber amplifiers (EDFA) (EDFA100S), one for each channel. Before amplifying the pulse, we use wavelength division multiplexers (WDM) (DWDM-SM-1-34-L-1-2) to couple a 1543 (nm) CW laser (DFB-1543-SM-30) to the pulses so that the spontaneous emission noise during the amplification is reduced. We decouple the 1550 (nm) pulses from the 1543 (nm) CW laser with the same WDM after the amplification is done. Finally, we use PC to ensure the correct linear polarization for the 1550 (nm) signal pulses. After this, a complete quantum walk step is finished.

\subsection{Mathematical formulation of the quantum walk experiment}
\begin{figure}[t]
%\subfloat[\label{sfig:1}]{}
\includegraphics[width=0.9\columnwidth]{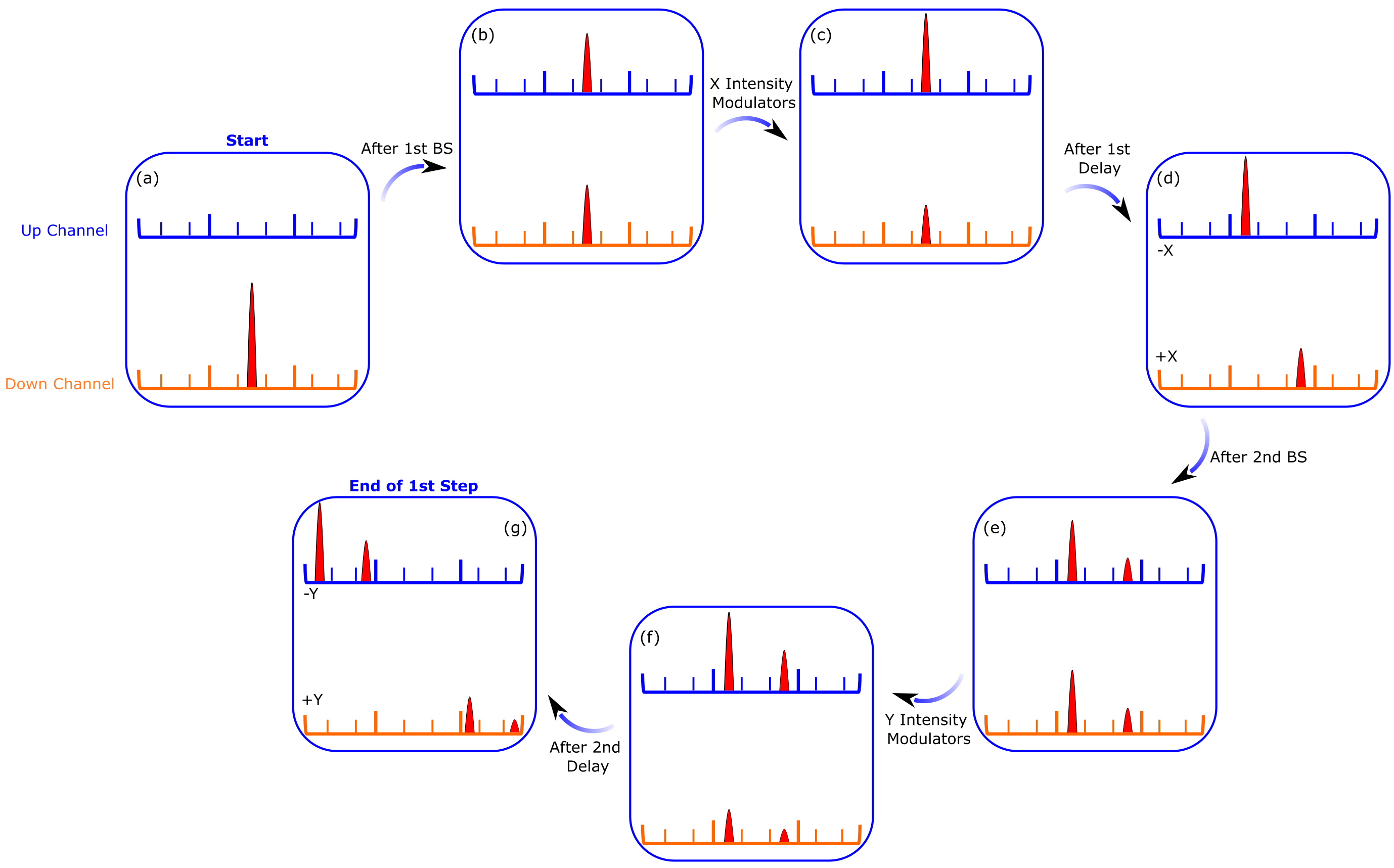}
\caption{\label{fig:s2} \textbf{Example of full evolution in a complete step.} Both the up and down channels of the fiber loops are shown. (a-g) The complete evolution within one step given that the initial state is $D_{x=0,y=0}=1$ and all other $U_{x,y}$ and $D_{x,y}$ are $0$.}
\end{figure}

Here we give the most general mathematical formulation of our quantum walk. Figure \ref{fig:s2}(a-g) shows this evolution when the input state is $D_{0,0}=1$.

As mentioned in the previous section, the state at each step is given by a complex vector $(U_{x,y},D_{x,y})$ where $x$ and $y$ ranges from $-15$ to $+15$. After the first beam splitter ($\pm X$), the state is updated to (Figure \ref{fig:s2}(b):
\begin{equation}
	\begin{aligned}
		U_{x,y}^{'} =  \frac{1}{\sqrt{2}}(U_{x,y}-D_{x,y})\\
		D_{x,y}^{'} =  \frac{1}{\sqrt{2}}(U_{x,y}+D_{x,y})
	\end{aligned} 
\end{equation}

After the first pair of modulators ($\pm X$ modulators), we obtain:
\begin{equation}
	\begin{aligned}
		U_{x,y}^{''} = U_{x,y}^{'}f^{(U)}_{x,y}\\
		D_{x,y}^{''} = D_{x,y}^{'}f^{(D)}_{x,y}
	\end{aligned}
\end{equation}
where $f^{(U/D)}_{x,y}$ is the gain/loss applied to each time bin in the up/down channel by the $X$ modulators.

After the delay:
\begin{equation}
	\begin{aligned}
		U_{x,y}^{'''} =  U_{x+1,y}^{''}\\
		D_{x,y}^{'''} =  D_{x-1,y}^{''}
	\end{aligned}
\end{equation}

This yields:
\begin{equation}
	\begin{aligned}
		U_{x,y}^{'''} =  \frac{1}{\sqrt{2}}(U_{x+1,y}-D_{x+1,y})f^{(U)}_{x+1,y}\\
		D_{x,y}^{'''} =  \frac{1}{\sqrt{2}}(U_{x-1,y}+D_{x-1,y})f^{(D)}_{x-1,y}\\
	\end{aligned}
\end{equation}

Then, we enter the second beamsplitter ($\pm Y$):
\begin{equation}
	\begin{aligned}
		W_{x,y}^{'} =  \frac{1}{\sqrt{2}}(U_{x,y}^{'''}-D_{x,y}^{'''})\\
		F_{x,y}^{'} =  \frac{1}{\sqrt{2}}(U_{x,y}^{'''}+D_{x,y}^{'''})
	\end{aligned} 
\end{equation}

After modulation ($\pm Y$ modulators):
\begin{equation}
	\begin{aligned}
		W_{x,y}^{''} = W_{x,y}^{'}c^{(U)}_{x,y}\\
		F_{x,y}^{''} = F_{x,y}^{'}c^{(D)}_{x,y}
	\end{aligned}
\end{equation}
where $c^{(U/D)}_{x,y}$ is the gain/loss applied to each time bin in the up/down channel by the $Y$ modulators.

After delay:
\begin{equation}
	\begin{aligned}
		W_{x,y}^{'''} = W_{x,y+1}^{''} = \frac{1}{\sqrt{2}}(U_{x,y+1}^{'''} - D_{x,y+1}^{'''})\times c^{(U)}_{x,y+1}\\
		F_{x,y}^{'''} =  F_{x,y-1}^{''} = \frac{1}{\sqrt{2}}(U_{x,y-1}^{'''} + D_{x,y-1}^{'''})\times c^{(D)}_{x,y-1}
	\end{aligned}
\end{equation}

$(W_{x,y}^{'''},F_{x,y}^{'''})$ is thus the output state given the input state $(U_{x,y},D_{x,y})$. Consider a pulse ending up in time bin $(x,y)$ at step $(n+1)$. Denoting it as $U^{(n+1)}_{x,y}$ and $D^{(n+1)}_{x,y}$, we have:
\begin{equation}
	\begin{aligned}
		U_{x,y}^{(n+1)} %&= \frac{1}{\sqrt{2}}(U_{x,y+1}^{(3)} - D_{x,y+1}^{(3)})c^{(U)}_{x,y+1}\\
		&= \frac{1}{2}[(U^{(n)}_{x+1,y+1} - D^{(n)}_{x+1,y+1})f^{(U)}_{x+1,y+1}c^{(U)}_{x+1,y+1} - 
		(U^{(n)}_{x-1,y+1} + D^{(n)}_{x-1,y+1})f^{(D)}_{x-1,y+1}c^{(U)}_{x-1,y+1}]\\
		D_{x,y}^{(n+1)} %&= \frac{1}{\sqrt{2}}(U_{x,y-1}^{(3)} + D_{x,y-1}^{(3)})c^{(D)}_{x,y-1}\\
		&= \frac{1}{2}[(U^{(n)}_{x+1,y-1} - D^{(n)}_{x+1,y-1})f^{(U)}_{x+1,y-1}c^{(D)}_{x+1,y-1}+
		(U^{(n)}_{x-1,y-1} + D^{(n)}_{x-1,y-1})f^{(D)}_{x-1,y-1}c^{(D)}_{x-1,y-1}]\\ 
	\end{aligned}
\end{equation}

The above equation thus describes the full evolution of the state within one step. We can obtain lattice gain/loss pattern as shown in Figure 1(a) or in Figure 1(c) of the main text, by properly choosing $f^{(U/D)}$ and $c^{(U/D)}$ as a function of $(x,y)$. For example, for a bulk lattice as shown in Figure 1(a) of the main text, we choose $f^{(U)}_{x,y} = \alpha e^{0.175}$, $f^{(D)}_{x,y} = \alpha e^{-0.175}$, $c^{(U)}_{x,y} = \alpha e^{0.175}$, $c^{(D)}_{x,y} = \alpha e^{-0.175}$. Here $\alpha$ is the additional loss imposed by the modulators, and we use EDFA to compensate for the loss such that effectively, $\alpha=1$.
In the experiment, we are only measuring light in the up channel, and thus the power of pulses, or the probability distribution of the walker in the up channel, is $P_{x,y}\propto|U_{x,y}|^2$. We normalize this probability distribution for all experiment data. 
%The functions $f^{(U/D)}_{x,y}$ and $c^{(U/D)}_{x,y}$ thus determins the gain/loss pattern in the 2D synthetic lattice. 
%For example, for a bulk lattice as shown in Figure 1(a) of the main text, we choose $f^{(U)}_{x,y} = \alpha e^{0.175}$, $f^{(D)}_{x,y} = \alpha e^{-0.175}$, $c^{(U)}_{x,y} = \alpha e^{0.175}$, $c^{(D)}_{x,y} = \alpha e^{-0.175}$. Here $\alpha$ is the additional loss imposed by the modulators, and we use EDFA to compensate for the loss such that effectively, $\alpha=1$.

\subsection{Effective band and energy winding of the bulk model}
\begin{figure}[ht]
\includegraphics[width=0.75\columnwidth]{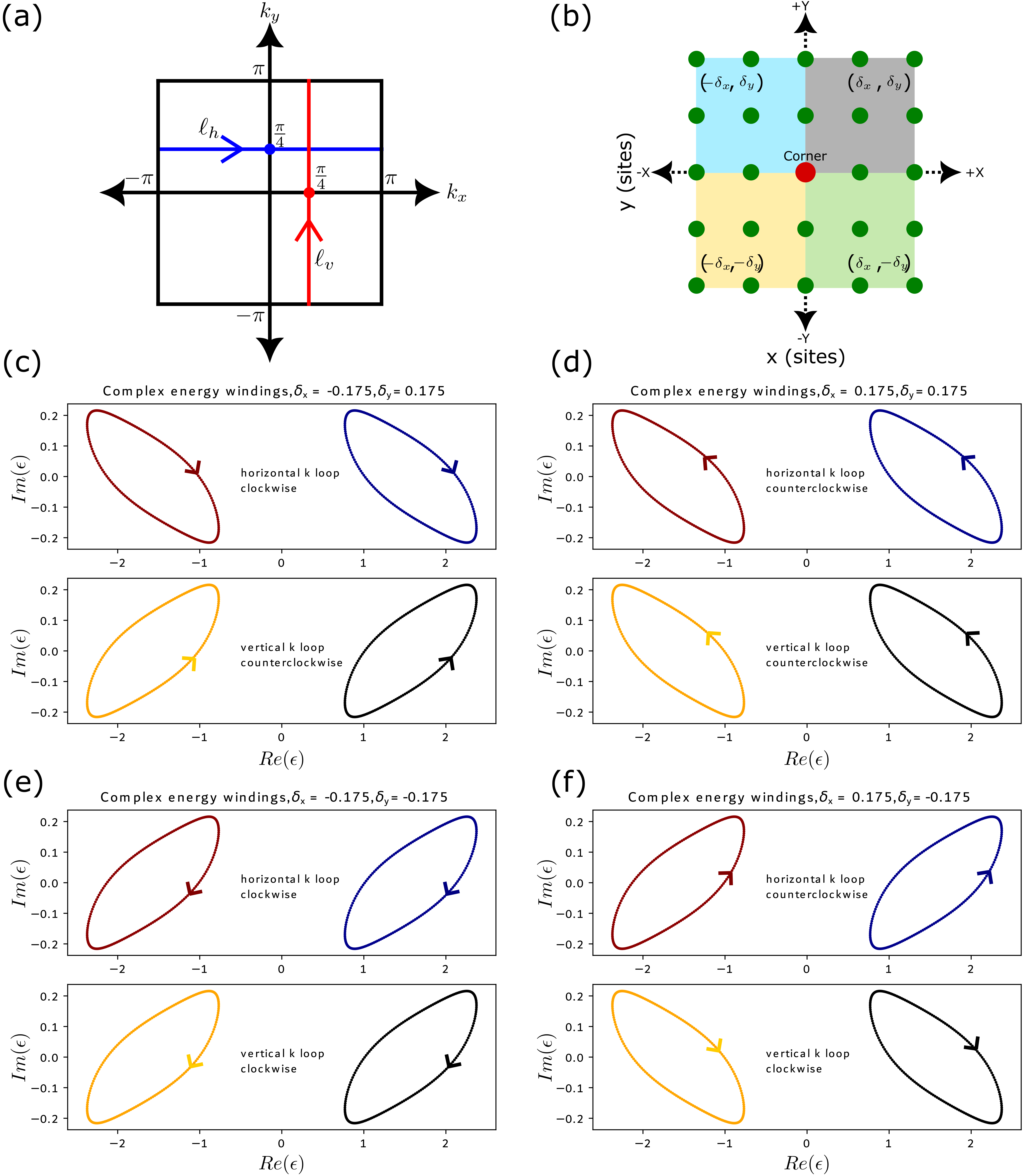}
\caption{\label{fig:s3} \textbf{Bulk band non-Hermitian topology.} (a) For each bulk patch, we choose two oriented loops in the Brillouin zone: $l_h = \{k_y\equiv \pi/4, k_x\in [-\pi\rightarrow \pi]\}$ and $l_v=\{k_x\equiv \pi/4, k_y\in [-\pi\rightarrow\pi]\}$. Each loop $l_i$ then contributes to two individual complex energy winding loops $\epsilon_{up}(l_i)$ and $\epsilon_{down}(l_i)$ winding in the same direction. We calculate this winding for all four patches in (b), corresponding to positive or negative $\delta_x$ and $\delta_y$. (c-f) The topological invariant is the winding direction of the directed curve $\epsilon(l_h)$ and $\epsilon(l_v)$ in the complex energy plane, which can either be clockwise or counterclockwise.}
\end{figure}

Here we impose periodic boundary conditions for the bulk lattice as shown in Figure 1(a) of the main text, in both $X$ and $Y$ directions. The modulations are $f^{(U)}_{x,y} = e^{0.175}$, $f^{(D)}_{x,y} = e^{-0.175}$, $c^{(U)}_{x,y} = e^{0.175}$, $c^{(D)}_{x,y} = e^{-0.175}$, and thus $\delta_x=0.175$ and $\delta_y=0.175$. With this assumption, we can therefore apply the Bloch theorem to the quantum walk evolution equation in the previous section and introduce the Bloch vector $(k_x,k_y)$. We use the ansatz $U_{x,y}=e^{ik_x x+ik_y y}\tilde{U}_{k_x,k_y}$ and $D_{x,y}=e^{ik_x x+ik_y y}\tilde{D}_{k_x,k_y}$ for the eigenmodes of the walk. The evolution equation can now be simplified to:

\begin{equation}
	\begin{aligned}
 \begin{bmatrix}
 \tilde{U}_{k_x,k_y} \\ \tilde{D}_{k_x,k_y}
 \end{bmatrix}
 =
\frac{1}{2}\begin{bmatrix}
e^{ik_x+ik_y}f^{(U)}c^{(U)} - e^{-ik_x+ik_y}f^{(D)}c^{(U)} && -e^{ik_x+ik_y}f^{(U)}c^{(U)} - e^{-ik_x+ik_y}f^{(D)}c^{(U)}\\ 
e^{ik_x-ik_y}f^{(U)}c^{(D)} + e^{-ik_x-ik_y}f^{(D)}c^{(D)} && -e^{ik_x-ik_y}f^{(U)}c^{(D)} + e^{-ik_x-ik_y}f^{(D)}c^{(D)}\\ 
\end{bmatrix} 
\begin{bmatrix}
\tilde{U}_{k_x,k_y} \\ \tilde{D}_{k_x,k_y}
\end{bmatrix}
	\end{aligned}
\end{equation}

Since we have two discrete degree of freedom $U$ and $D$, we always obtain two different eigenvalues for each $(k_x,k_y)$ as we diagonalize the above $2\times2$ matrix. We call the two eigenvalues $u_{k_x,k_y}$ and $d_{k_x,k_y}$. The two effective energies are thus defined as $\epsilon_{up}(k_x,k_y) = ilog(u_{k_x,k_y})$ and $\epsilon_{down}(k_x,k_y) = ilog(d_{k_x,k_y})$.

We further consider the four bulk patches in Figure \ref{fig:s3}(b) and show that they exhibit different non-Hermitian topological invariants, namely the winding of the effective energy $\epsilon_{up/down}(k_x,k_y)$ in the complex plane. Without loss of generality, we always pick two loops in the Brillouin zone: $l_v = \{k_x\equiv \pi/4, k_y\in [-\pi\rightarrow \pi]\}$ and $l_h=\{k_y\equiv \pi/4, k_x\in [-\pi\rightarrow\pi]\}$, as shown in Figure \ref{fig:s3}(a). For each bulk panel in Figure \ref{fig:s3}(b), as one varies $(k_x,k_y)$ along $l_v$ and $l_h$, the corresponding complex energy $\epsilon(k_x,k_y)$ can finish a single loop in the complex plane, either in the clockwise or counterclockwise direction. The winding direction of $\epsilon(l_v)$ and $\epsilon(l_h)$ forms the non-Hermitian topological invariant of the bulk. Note that we have suppressed the unimportant label $up$ and $down$ since $\epsilon_{up}(l_i)$ and $\epsilon_{down}(l_i)$ always wind in the same direction.

\subsection{Localized eigenmodes at the presence of boundary and corner}

\begin{figure}[t]
\includegraphics[width=0.9\columnwidth]{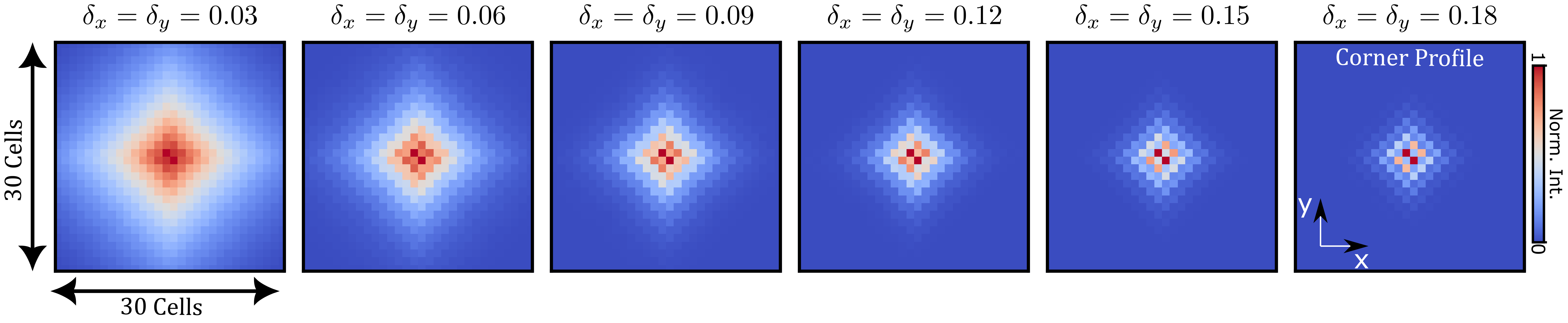}
\caption{\label{fig:s4} \textbf{Averaged eigenmode spatial profile for different non-Hermitian parameters $\delta_x=\delta_y$.} We adopt the lattice geometry as in Figure \ref{fig:s2}(b). From left to right we take values $0.03,0.06,0.09,0.12,0.15$ and $0.18$.}
\end{figure}

As shown in Figure 1(d) in the main text, the averaged spatial profile of the eigenmodes of the walk is localized at the corner. In Figure 1(c) of the main text we have chosen $\delta_x=\delta_y=0.175$, but the feature of the spatial profile persists for any $\delta_x=\delta_y>0$. Here in Figure \ref{fig:s4} of the supplementary section, we show the average eigenmode spatial profile for $\delta_x=\delta_y=0.03,0.06,0.09,0.12,0.15$ and $0.18$. As one increases $\delta_x=\delta_y$, we observe that the averaged spatial profile becomes more localized. This explains the adiabatic tapering of the walker's probability distribution as one increases/decreases the non-Hermitian parameter in time, as shown in Figure 3(a) in the main text.

\subsection{Static control supplementary data}

\begin{figure}[t]
\includegraphics[width=0.9\columnwidth]{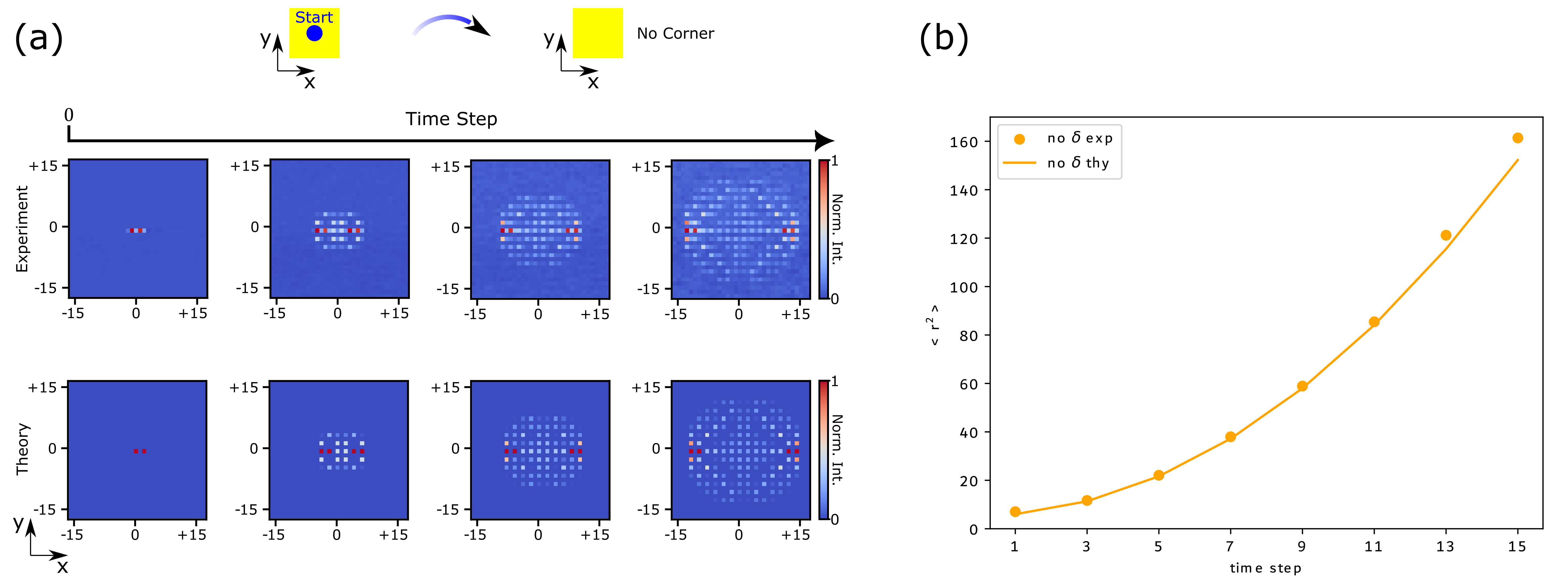}
\caption{\label{fig:s5} \textbf{Evolution of probability distribution for non-Hermitian parameters $\delta_x=\delta_y=0$, showing diffusive spreading.} (a) Probability distribution evolution, where the snapshots are taken at time step $1,5,9,13$, respectively. (b) Evolution of averaged displacement $<r^2>=<x^2+y^2>=\sum_{x,y} P_{x,y}(x^2+y^2)$ for time step from $1,3,5,7,9,11,13$ and $15$.}
\end{figure}
Here we present additional experimental results for the quantum walk with no dynamical control. We first show that, with $\delta_x=\delta_y=0$, the walker diffusively spreads into the bulk of the lattice, as shown in Figure \ref{fig:s5}(a), where the probability distribution of the walker is plotted for step $1,5,9,13$. The averaged displacement, defined as $<r^2>(n) = \sum_{x,y} P_{x,y}(n)(x^2+y^2)$, is plotted in Figure \ref{fig:s5}(b), for step $1,3,5,7,9,11,13$ and $15$. Here $P_{x,y}$ is the probability distribution of the walker at time step $n$. 

\begin{figure}[t]
\includegraphics[width=0.9\columnwidth]{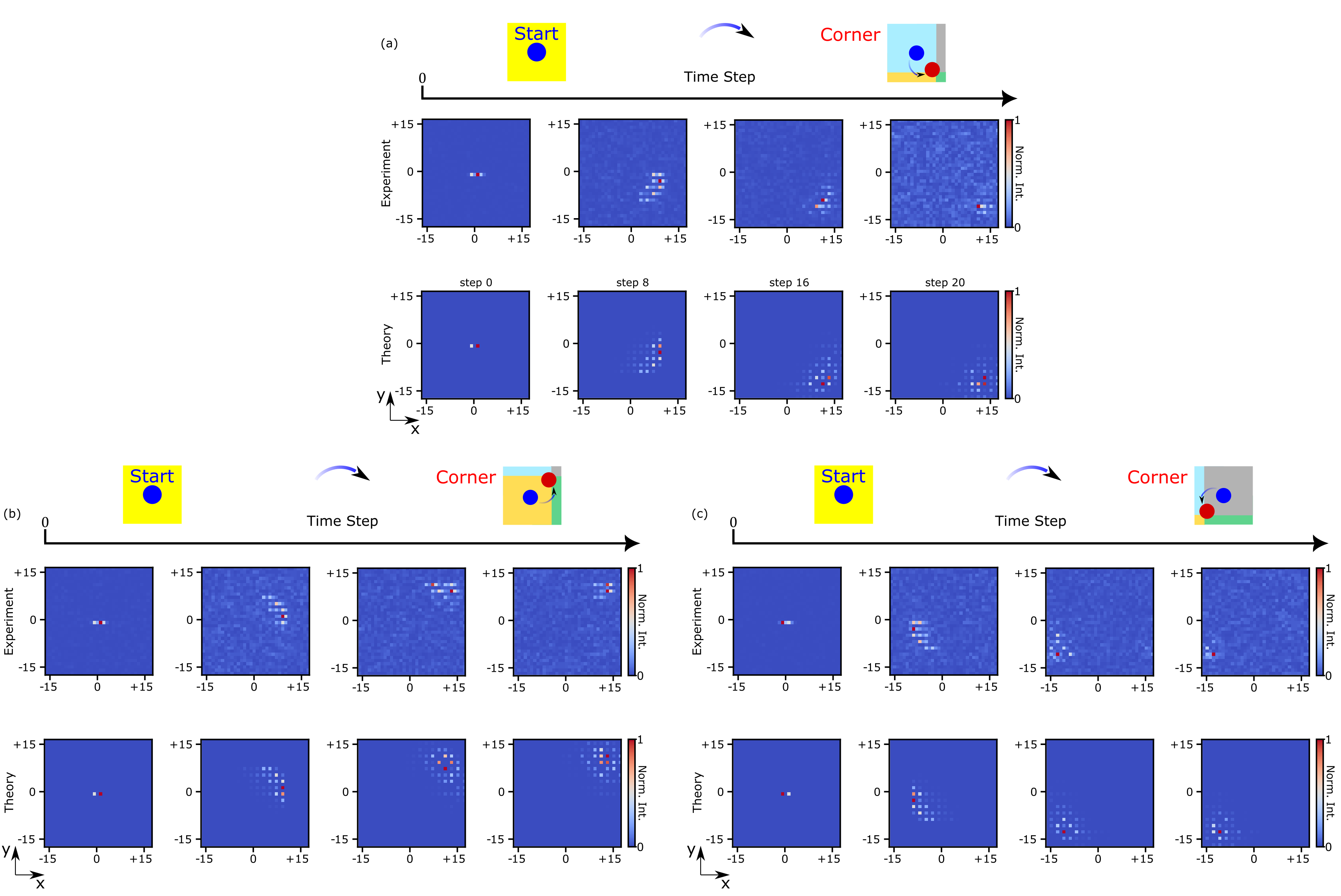}
\caption{\label{fig:s6} \textbf{Funneling and stabilization of light starting from arbitrary bulk patches.} Light is always initialized at $(x,y)=(0,0)$, but the corner position is held fixed at (a) $(10,-10)$, (b) $(10,10)$ and (c) $(-10,-10)$. For each panel, the top row is probability distributions collected at step $1,9,17,21$ of the experiment, and the bottom row is the corresponding simulation results.}
\end{figure}

Furthermore, as mentioned in the main text, the funneling of light happens wherever the walker is initialized, assuming the lattice gain-loss pattern shown in Figure 1(c) of the main text. This is manifestly shown in Figure \ref{fig:s6}, where we always choose the initial state to be $D_{x=0,y=0}=1$, but lattice corner is located at $(10,-10)$, $(10,10)$ and $(-10,-10)$, respectively.

%%%%%%%%%%%%%%%%%%%%%%%%%%%%%%%%%%%%%%%%%%%%%%%%%%%%%%%%%%%%%%%%%%%%%%%%%%%%%%%%%%%%%%%%%%%%%%%%%%%%%%%%%%%%%%%%%%%%%%%%%%%%%%
\end{document}